\documentclass[twocolumn,twoside,slac_two]{revtex4}
\usepackage{graphicx}
\usepackage{fancyhdr}
\begin{document}

\title{TANAMI - Tracking Active Galactic Nuclei with Austral Milliarcsecond Interferometry}

\author{\mbox{Cornelia M\"uller$^{1,2}$}, \mbox{Matthias Kadler$^{1,2,3,4}$}, \mbox{Roopesh Ojha$^{5,6}$}, \mbox{M. B\"ock$^{1,2}$}, \mbox{R. Booth$^{7}$}, \mbox{M. S. Dutka$^{8}$},\mbox{ P. G. Edwards$^{9}$}, \mbox{A. L. Fey$^{5}$}, \mbox{L. Fuhrmann$^{10}$}, \mbox{H. Hase$^{11}$},\mbox{ S. Horiuchi$^{12}$}, \mbox{F. Hungwe$^{13}$}, \mbox{D. L. Jauncey$^{9}$}, \mbox{K. J. Johnston$^{5}$}, \mbox{U. Katz$^{2}$}, \mbox{M. Lister$^{13}$}, \mbox{J. E. J. Lovell$^{14}$}, \mbox{C. Pl\"otz$^{15}$}, \mbox{J. F. H. Quick$^{7}$}, \mbox{E. Ros$^{10
,17}$}, \mbox{G. B. Taylor$^{18}$}, \mbox{D. J. Thompson$^{19}$}, \mbox{S. J. Tingay$^{20}$}, \mbox{G. Tosti$^{21,22}$}, \mbox{A. K. Tzioumis$^{9}$}, \mbox{J. Wilms$^{1,2}$}, \mbox{J. A. Zensus$^{10}$}}
\affiliation{
$^{1}$\,Dr. Remeis-Sternwarte \& ECAP, Sternwartstr.~7, 96049 Bamberg, Germany\\
$^{2}$\,ECAP, Universit\"at Erlangen-N\"urnberg, Germany\\
$^{3}$\,CRESST/NASA Goddard Space Flight Center, Greenbelt, MD 20771, USA\\
$^{4}$\,USRA, 10211 Wincopin Circle, Suite 500 Columbia, MD 21044, USA\\
$^{5}$\,United States Naval Observatory, 3450 Massachusetts Ave., NW, Washington DC 20392, USA\\
$^{6}$\,NVI,Inc., 7257D Hanover Parkway, Greenbelt, MD 20770, USA\\
$^{7}$\,Hartebeesthoek Radio Astronomy Observatory, PO Box 443, Krugersdorp 1740, South Africa\\
$^{8}$\,The Catholic University of America, 620 Michigan Ave., N.E.,  Washington, DC 20064, USA\\
$^{9}$\,CSIRO Astronomy and Space Science, PO Box 76, Epping, NSW 1710 Australia\\
$^{10}$\,MPIfR, Auf dem H\"ugel 69, 53121 Bonn, Germany\\
$^{11}$\,BKG, Univ. de Concepcion, Casilla 4036, Correo 3, Chile\\
$^{12}$\,Canberra Deep Space Communication Complex, PO Box 1035, Tuggeranong, ACT 2901,  Australia\\
$^{13}$\,Rhodes University, P.O. Box 94 Grahamstown 6140, South Africa\\
$^{14}$\,Dept. of Physics, Purdue University, 525 Northwestern Avenue, West Lafayette, IN 47907, USA\\
$^{15}$\,School of Mathematics \& Physics, Private Bag 37, Univ. of Tasmania, Hobart TAS 7001, Australia\\
$^{16}$\,BKG, Geodetic Observatory Wettzell, Sackenrieder Str. 25, 93444 Bad K\"otzting, Germany\\
$^{17}$\,Dept. d'Astronomia i Astrof\'{\i}sica, Universitat de Val\`encia, 46100 Burjassot, Val\`encia, Spain\\
$^{18}$\,Dept. of Physics and Astronomy, University of New Mexico, Albuquerque NM, 87131, USA\\
$^{19}$\,Astrophysics Science Division, NASA Goddard Space Flight Center, Greenbelt, MD 20771, USA\\
$^{20}$\,Curtin Institute of Radio Astronomy, Curtin University of Technology, Bentley, WA, 6102, Australia\\
$^{21}$\,Istituto Nazionale di Fisica Nucleare, Sezione di Perugia, 06123 Perugia, Italy\\
$^{22}$\,Dipartimento di Fisica, Universit\`a degli Studi di Perugia, 06123 Perugia, Italy\\
}
\begin{abstract}
We present a summary of the observation strategy of TANAMI (Tracking Active Galactic Nuclei with Austral Milliarcsecond Interferometry), a monitoring program to study the parsec-scale structure and dynamics of relativistic jets in active galactic nuclei (AGN) of the Southern Hemisphere with the Australian Long Baseline Array (LBA) and the trans-oceanic antennas Hartebeesthoek, TIGO, and O'Higgins. TANAMI is focusing on extragalactic sources south of $\delta=-30^\circ$ with observations at $8.4\,$GHz and $22\,$GHz every $\sim$2 months at milliarcsecond resolution. The initial TANAMI sample of 43 sources has been defined before the launch of the \textsl{Fermi Gamma Ray Space Telescope} to include the most promising candidates for bright $\gamma$-ray emission to be detected with its Large Area Telescope (LAT). Since November 2008, we have been adding new sources to the sample, which now includes all known radio- and $\gamma$-ray bright AGN of the Southern Hemisphere. The combination of VLBI and $\gamma$-ray observations is crucial to understand the broadband emission characteristics of AGN and the nature of relativistic jets. 
\end{abstract}

\maketitle

\thispagestyle{fancy}

\section{Introduction}
Blazars are a radio-loud sub-group of Active Galactic Nuclei (AGN) emitting light across the whole electromagnetic spectrum. Former studies of the broadband luminosity variations suggested a close connection between radio and $\gamma$-ray emission, spectral changes, outbursts and the ejection of parsec-scale jet components~\cite{Dondi1995}. Blazars typically show superluminal motion of discrete jet features. Observations indicate that blazars are very compact and have a parsec-scale radio jet pointing close to our line of sight. To answer fundamental questions concerning beaming factors of superluminal motion, origin of high energy production, and the connection between $\gamma$-ray flares and jet-component ejections, multiwavelength observations are crucial.

Very Long Baseline Interferometry (VLBI) provides high resolution images of the innermost parsec-scale structure of extragalactic jets allowing us to determine crucial jet parameters like apparent jet speed, Doppler factors, and opening and inclination angles.
TANAMI (Tracking Active Galactic Nuclei with Austral Milliarcsecond Interferometry) is a radio monitoring program which studies on the parsec-scale structure of Blazar jets south of \mbox{$-30^\circ$ declination}. The observations are made with the antennas of the Australian Long Baseline Array (LBA) and associated telescopes in the Southern Hemisphere (see Sect.~\ref{sec:Obs}). The initial sample of 43 sources is monitored approximately every two months at two frequencies, $8.4\,$GHz (X-band) and $22\,$GHz (K-band). TANAMI is the complement to other ongoing VLBI monitoring programs in the Northern Hemisphere (e.g. MOJAVE,~\cite{Lister2009}) and offers an invaluable data set of AGN in the Southern sky.

The inital TANAMI sample (see~Sect.~\ref{sec:sample}) is a radio selected flux-density limited subsample and a $\gamma$-ray selected subsample of known and candidate $\gamma$-ray sources based on results of \textsl{CGRO}/EGRET. The $\gamma$-ray observatory \textsl{Fermi} was the immediate driver for this multiwavelength project and therefore TANAMI started in November 2007 before the launch of \textsl{Fermi} to determine a proper observation cadence for all sources.
Here, we discuss the TANAMI sample selection criteria (see~Sect.~\ref{sec:sample}), describe the observations and data reduction procedures (see~Sect.~\ref{sec:Obs}) and present examples of first epoch images at 8.4\,GHz and 22\,GHz (Ojha et al., 2010, submitted to A\&A).
%%%%%%%%%%%%%%%%%%%%%
\section{Sample Selection}\label{sec:sample}
The initial TANAMI sample consists of 43 blazars located south of $-30^\circ$ declination. It has been defined combining a radio selected flux-density limited subsample and a $\gamma$-ray selected subsample of known and candidate $\gamma$-ray sources based on results of \textsl{CGRO}/EGRET~\cite{Hartman1992}. We included all known radio sources south of $\delta=-30^\circ$ from the catalogue of Stickel et al.~\cite{Stickel1994} which are above a limiting radio flux density of \mbox{$S_{5\,\text{GHz}}>2\,$Jy}, and which have a flat radio spectrum \mbox{($\alpha>-0.5$, $S\sim\nu^{+\alpha}$)} between $2.7\,$GHz and $5\,$GHz. The $\gamma$-ray selected subsample includes all known $\gamma$-ray blazars detected by EGRET south of $\delta=-30^\circ$. Our sample-selection criteria proved very efficient in including $\gamma$-ray bright AGN jets. After three months of observations \textsl{Fermi} found 12 of our inital 43 source sample ($28\%$) to be bright $\gamma$-ray sources with $> 10 \sigma$ significance.

Furthermore we have been monitoring new $\gamma$-ray bright AGN from the \textsl{Fermi} LAT Bright AGN Sample (LBAS)~\cite{Abdo2009} list since November 2008. During 2009, we continued adding newly detected non-LBAS \textsl{Fermi} sources so that by now our sample is comprised of 75 sources, thereof are 10 recently added from the preliminary version of First LAT Catalog (Abdo et al, in preparation). Table\,I contains the source list of the enlarged sample.
%%%%%%%%%%%
\begin{table*}
\begin{center}
\caption{Source list of extended TANAMI sample}
\begin{tabular}[t]{ccccc}\hline
IAU Name & Alt Name & LAT$^{a}$ & ID$^{b}$ & $z$\\\hline
0047$-$579 & 	& 			Y & Q & 1.797\\
0055$-$328	&		&	Y	&	U	&	-	\\
0208$-$512 &	& 			\textbf{Y} & B & 0.99\\
0227$-$369	&		&	\textbf{Y}	&	Q	&	2.115	\\
0244$-$470	&		&	\textbf{Y}	&	U	&	-	\\
0302$-$623	&		&	Y	&	Q	&	1.351	\\
\textit{0308$-$611} &	&	Y	&	Q & - \\
0332$-$376	&		&	Y	&	U	&	-	\\
0332$-$403 &	& 			\textbf{Y} & B & 1.445\\
\textit{0402$-$362} &	&	Y & Q & 1.42 \\
0405$-$385 &	& 			\textbf{Y} & Q & 1.285\\
0412$-$536	&		&	\textbf{Y}	&	U	&	-	\\
0426$-$380	&		&	Y	&	Q	&	1.112	\\
0438$-$436&	& 			N & Q & 2.863\\
0447$-$439	&		&	\textbf{Y}	&	B	&	0.107	\\
0454$-$463 & 	& 			Y & Q & 0.853\\
0506$-$612&	& 			Y & Q & 1.093\\
0516$-$621	&		&	\textbf{Y}	&	U	&	-	\\
0518$-$458& PICTOR A & 		N & G & 0.0351\\
0521$-$365 & ESO 362$-$G021& 	Y & B & 0.0553\\
0524$-$485	&		&	Y	&	U	&	-	\\
0527$-$359	&		&	N	&	U	&	-	\\
0537$-$441& 	& 			\textbf{Y} & Q & 0.894\\
0625$-$354 & OH$-$342 & 		Y & G & 0.0546\\
0637$-$752 &	& 			Y & Q & 0.653\\
0700$-$661	&		&	\textbf{Y}	&	U	&	-	\\
0717$-$432	&		&	Y	&	U	&	-	\\
\textit{0736$-$770} &	& Y &	Q & - \\
\textit{0745$-$330} &	& Y & U &  - \\
0812$-$736	&		&	Y	&	U	&	-	\\
1057$-$797	&		&	\textbf{Y}	&	U	&	-	\\
\textit{1101$-$536} &	& Y  & Q & -\\
1104$-$445&	&			N & Q & 1.598\\
1144$-$379 &	&			\textbf{Y} & Q & 1.048\\
1257$-$326 &	&			N & Q & 1.256\\
1258$-$321& ESO 443- G 024	&	Y	&	G	&	0.017	\\
1313$-$333&	&			Y & Q & 1.21\\
1322$-$428& CenA, NGC 5128 &	\textbf{Y} & G & 0.002\\
\hline
\end{tabular}
\hspace{0.08\linewidth}
\begin{tabular}[t]{ccccc}\hline
IAU Name & Alt Name & LAT$^{a}$ & ID$^{b}$ & $z$\\\hline
1323$-$526 & &				Y & U & -\\
\textit{1325$-$558} &	&	Y & U& - \\
1333$-$337 & IC4296 & 		N & G & 0.012\\
1344$-$376	&		&	Y	&	U	&	-	\\
1424$-$418 &	&			Y & Q & 1.522\\
1440$-$389	&		&	Y	&	U	&	0.066	\\
1454$-$354 &	&			\textbf{Y} & Q & 1.424\\
1501$-$343 &   & 			Y & U & - \\
1505$-$496	&		&	N	&	U	&	-	\\
1549$-$790 & 	& 			N & G & 0.150\\
\textit{1600$-$445} &	&	Y	&	U &	\\
\textit{1600$-$489} &	&	Y	&	U &	\\
1606$-$667	&		&	Y	&	U	&	-	\\
1610$-$771&	&			Y & Q & 1.71\\
\textit{1613$-$586} &	&	Y	&	Q &	\\
\textit{1646$-$506} &	&	Y	&	U&	\\
1714$-$336&	&			\textbf{Y} & B & - \\
1716$-$771&  &		N & U & - \\
1718$-$649 &	NGC 6328 &			N & G & 0.014\\
1733$-$565 &	&			N & G & 0.098\\
1759$-$396&	&			\textbf{Y} & Q & 0.296\\
1804$-$502&  &				N & Q & 1.606\\
1814$-$637&	&			N & G & 0.063\\
1933$-$400& 	&			Y & Q & 0.965\\
1934$-$638&	&			N&G	& 0.18  \\
1954$-$388&	&			Y* & Q & 0.63\\
2005$-$489 &	&			\textbf{Y} & B & 0.071\\
2027$-$308 & ESO 462$-$G 027 &N & G & 0.02\\
2052$-$474&	&		\textbf{Y} & Q & 1.489\\
2106$-$413&	&N & Q & 1.058\\
2136$-$428	&		&	\textbf{Y}	&	B	&	-	\\
2149$-$306&	&N & Q & 2.345\\
2152$-$699& ESO 075$-$G 041&N & G & 0.028\\
2155$-$304&	&\textbf{Y} & B & 0.116\\
2204$-$540&	&\textbf{Y}& Q & 1.206\\
2326$-$477&	&N & Q & 1.299\\
2355$-$534&	&N & Q & 1.006\\
\hline
\end{tabular}
% \label{table1}
\end{center}
\begin{flushleft}
\small{$^{a}$ based on the preliminary version of First LAT Catalog (Abdo et al, in preparation)\\
\small{\textbf{bold} denotes sources of the LAT 3-months list~\cite{Abdo2009}}\\
\small{\textit{italic} denotes all sources recently added to the TANAMI sample}\\
(*) denotes a low confidence detection}\\
\small{$^{b}$ optical classification, denoted according to~\cite{Veron2006} and~\cite{Healey2008} for all sources not appearing in~\cite{Veron2006} respectively, as:\\(Q) quasar, (B) BL Lac object, (G) galaxy, (U) unclassified}
\end{flushleft}
\end{table*}
%%%%%%%%%%%%%%%%%%%%%%%%
\section{Observations and Data Reduction}\label{sec:Obs}
\begin{figure}
\includegraphics[width=75mm]{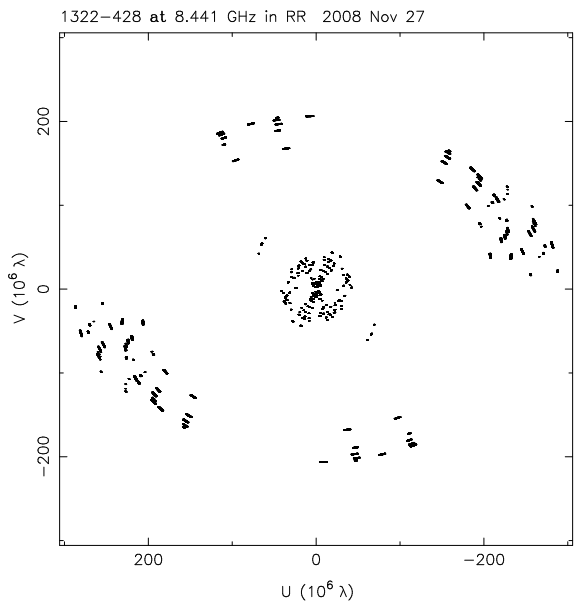}
\caption{Typical 8.4\,GHz ($u$-$v$)-coverage of Centaurus\,A}
\label{uv}
\end{figure}

TANAMI observations are made with a sub-set of the Australian Long Baseline Array (LBA), using the Australian antennas in Narrabri ($5\times22$\,m), Ceduna ($30\,$m), Hobart ($26\,$m), Mopra ($22\,$m), Parkes ($64\,$m), the $70\,$m and $34\,$m telescopes at NASA’s Deep Space Network (DSN) located at Tidbinbilla, and the $26\,$m South-African Hartebeeshoeck antenna (until 2008; currently inoperative). Since 2009, the $9\,$m German 
Antarctic Receiving Station (GARS) in O'Higgins, Antarctica, and the $6\,$m Transportable Integrated Geodetic Observatory (TIGO) in Chile participate in the program (at 8.4\,GHz, only). A typical ($u$--$v$)-coverage at 8.4\,GHz for a source at $-43^\circ$ declination is shown in Fig.~\ref{uv}.
The typical angular resolution achieved is about 0.6--0.9\,mas. The shorter baselines at the center of the plot are those between telescopes within Australia, the external longer baselines are those to TIGO and O'Higgins. Although intermediate-length baselines are absent, which limits the image quality, past imaging programs found that this constraint does not preclude good images as long as special care is taken in calibration and the imaging process (e.g.~\cite{Ojha2005}).

The data are recorded on the LBADRs (Long Baseline Array Disk Recorders) and correlated on the DiFX software correlator at Curtin University in Perth, Western Australia~\cite{Deller2007}. The correlated data were loaded into AIPS (National Radio Astronomy Observatory's Astronomical Image Processing System software) for data inspection, initial editing and fringe fitting. Amplitude Calibration was done by using known flux values of prior observed sources. The imaging was performed applying standard methods applying the program  {\sc difmap}~\cite{Shepherd1997}, using the {\sc clean} algorithm, giving the same weight to all visibility data points (natural weighting) and making use of phase self-calibration.

In general the best image fidelity is achieved at 8.4\,GHz. Observations at this frequency yield good resolved images registering almost all of the extended structure and hence we get here the most detailed information of the innermost jet features. Images at 22\,GHz show the jet morphology closer to the core. These dual-frequency observations offer spectral information which are an important component of broadband SEDs.
\begin{figure}
\includegraphics[width=75mm, clip]{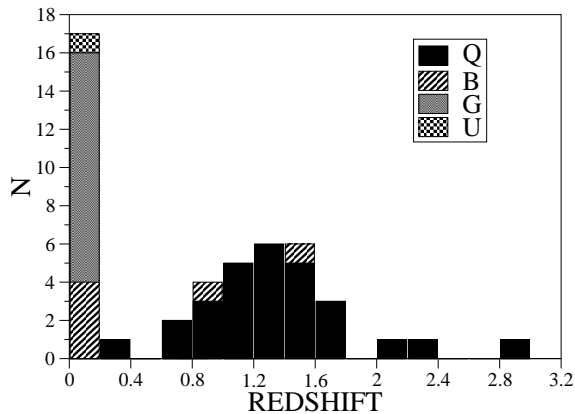}
\caption{Redshift distribution of the extended TANAMI sample for Quasars (Q), BL\,Lacs (B) \& Galaxies (G) and unclassified (U)}
\label{z}
\end{figure}
\begin{figure*}[htbp]
\includegraphics[width=0.28\textwidth]{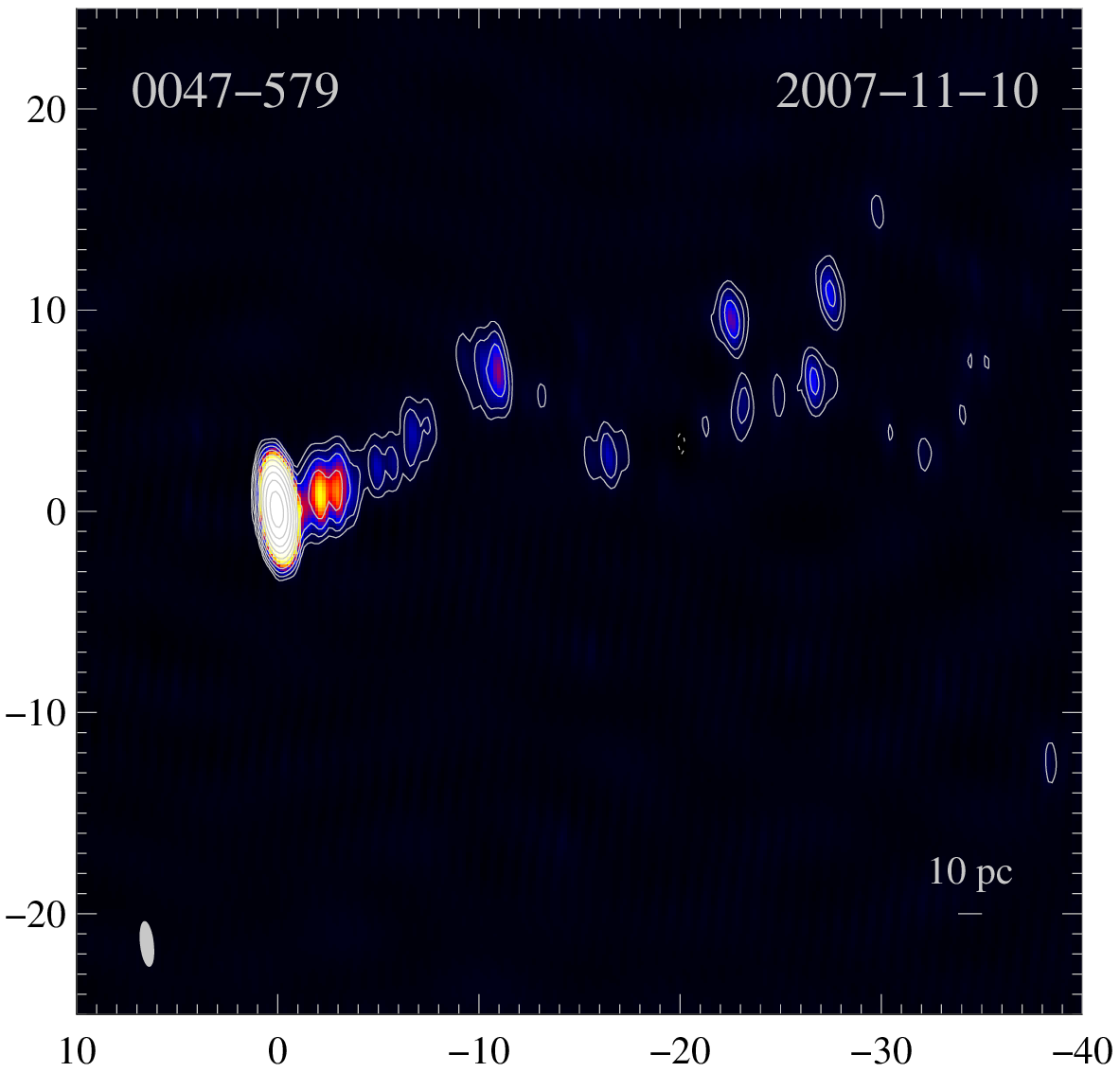}
\includegraphics[width=0.28\textwidth]{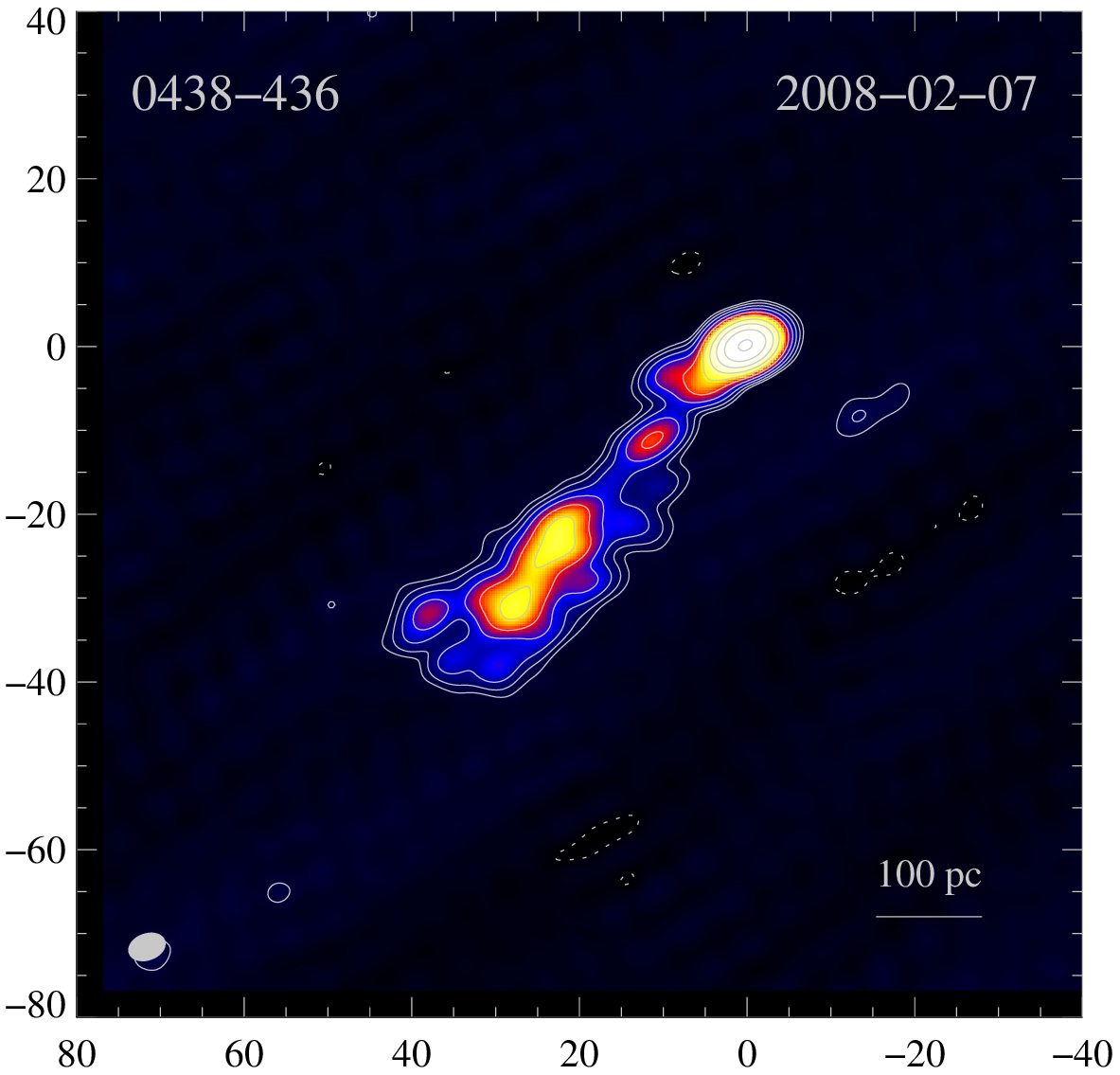}
\includegraphics[width=0.28\textwidth]{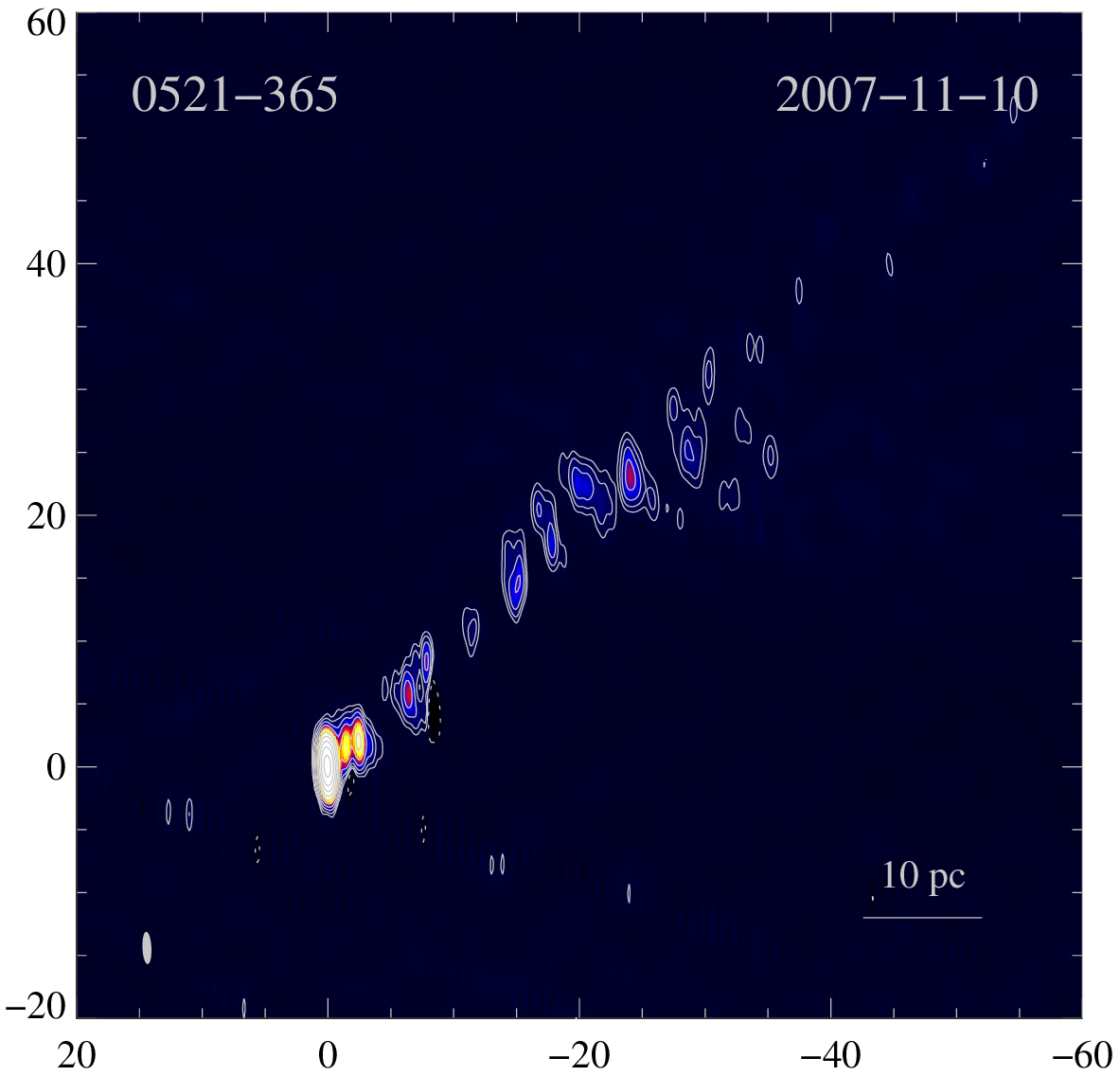}
\includegraphics[width=0.28\textwidth]{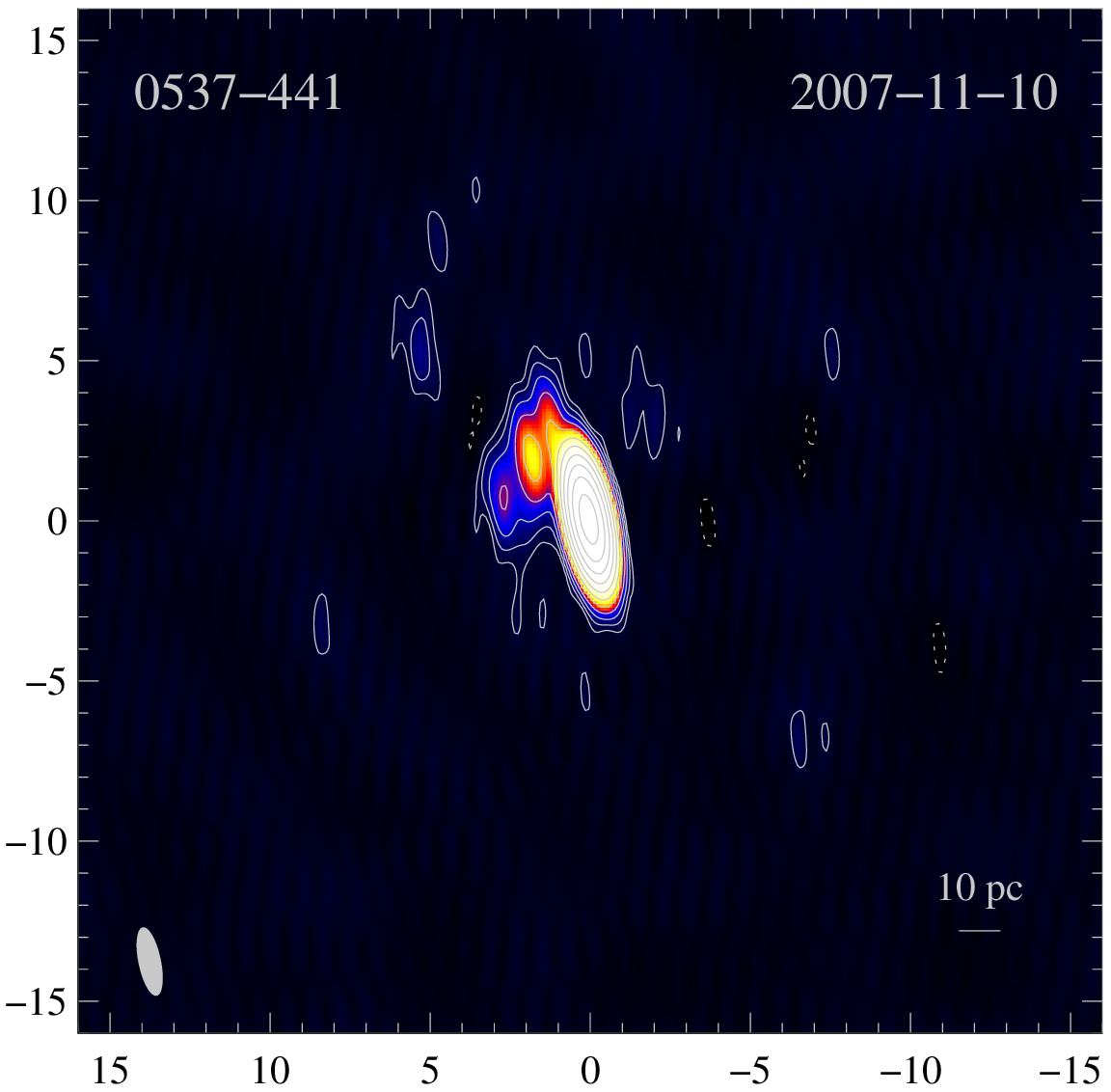}
\includegraphics[width=0.28\textwidth]{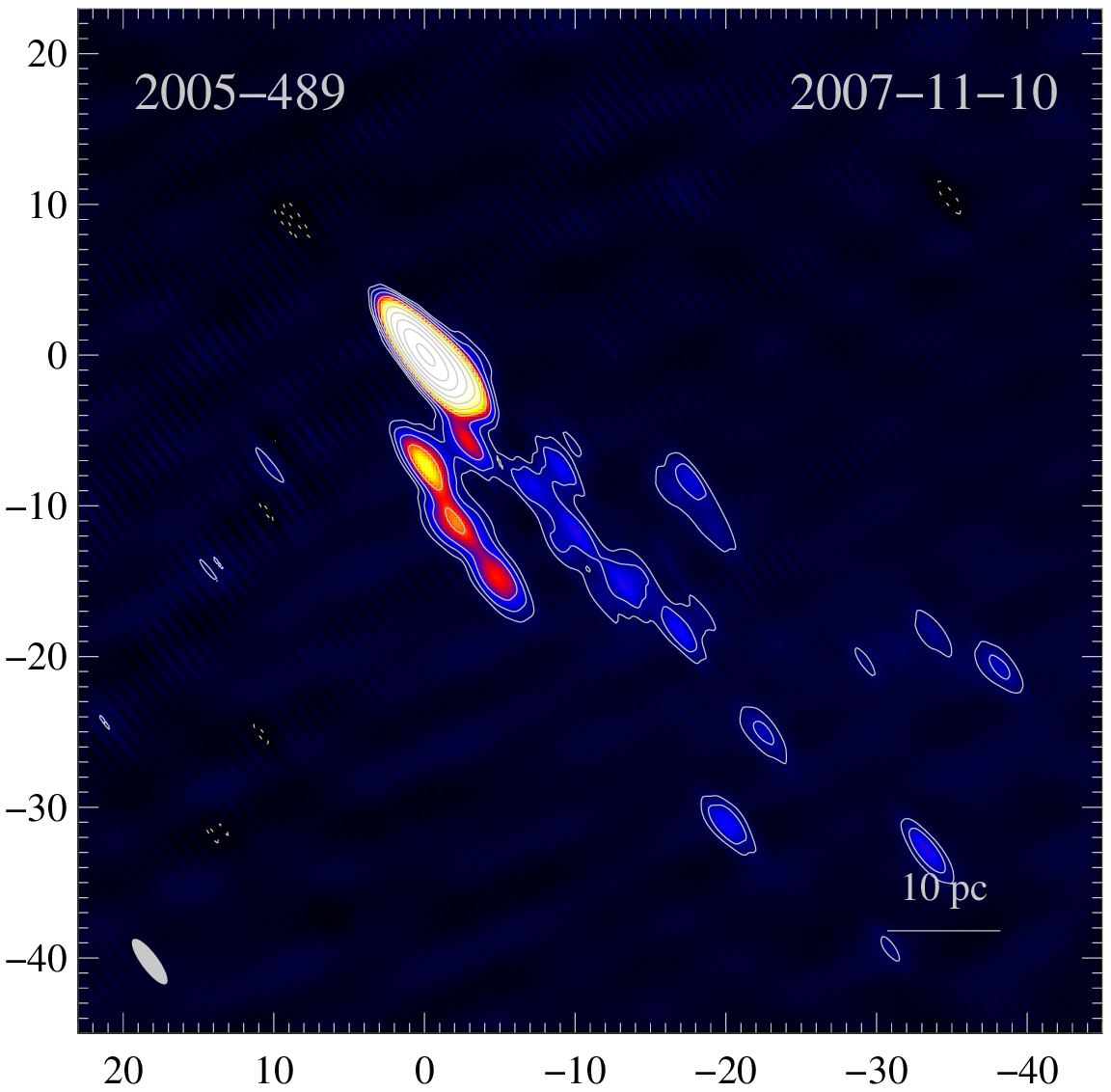}
\includegraphics[width=0.28\textwidth]{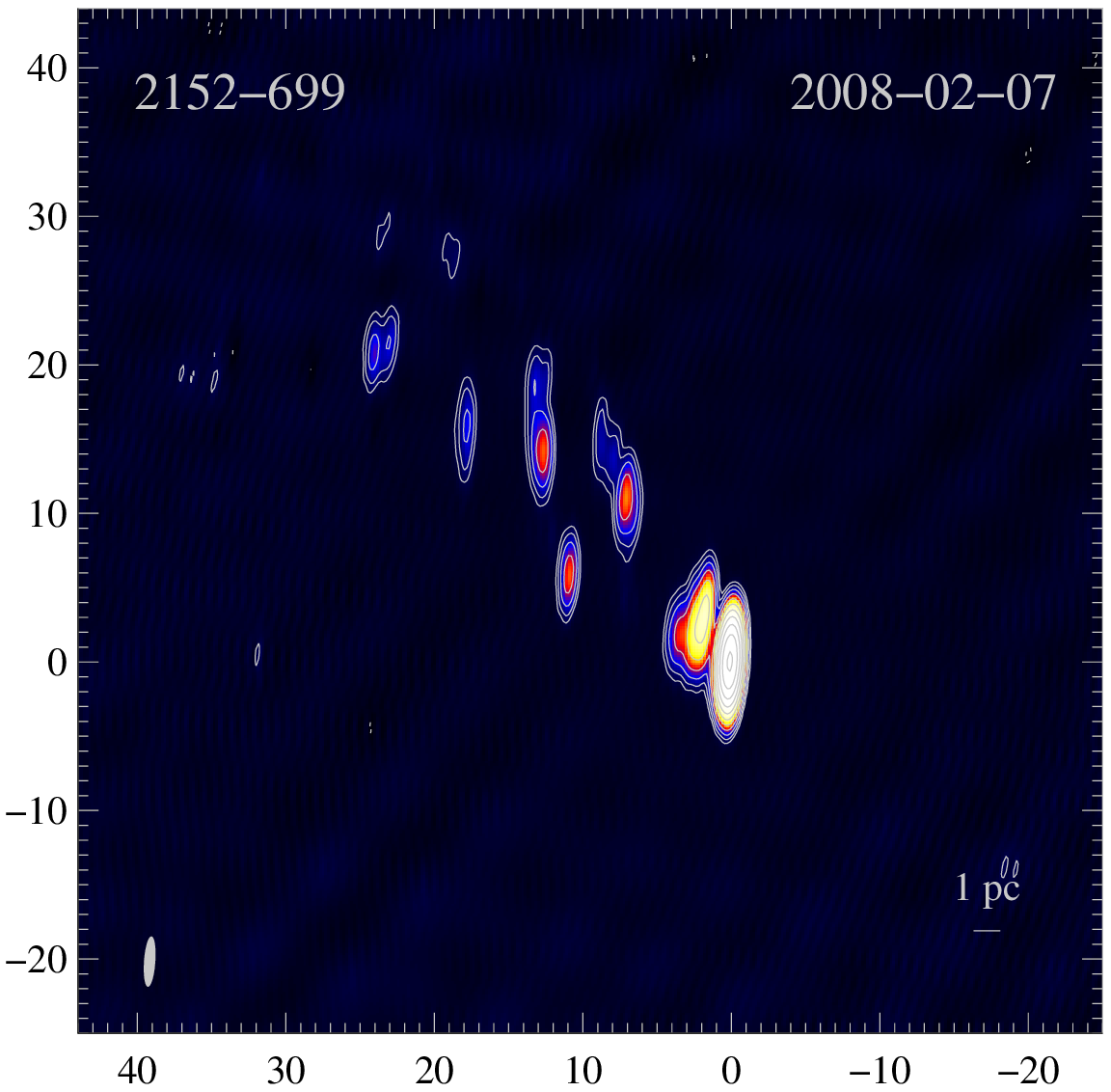}
\caption{Selection of TANAMI 8.4\,GHz images. Axis units are mas.}
\label{plots_X}
\end{figure*}
%%%%%%%%%%%%%%%%%%%%%%%%%%%%%%%%%5
\begin{figure*}[htbp]
\includegraphics[width=0.28\textwidth]{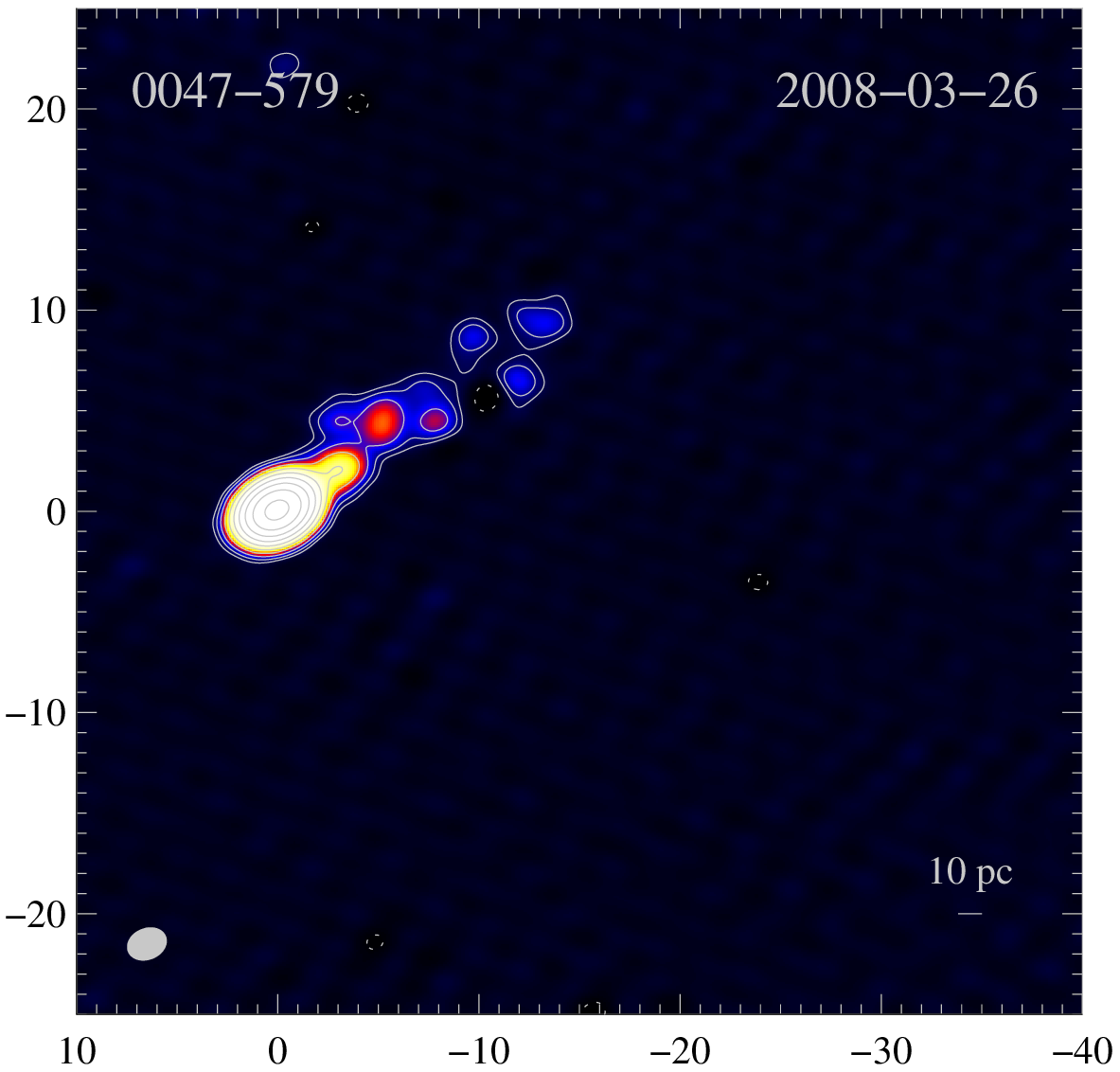}
\includegraphics[width=0.28\textwidth]{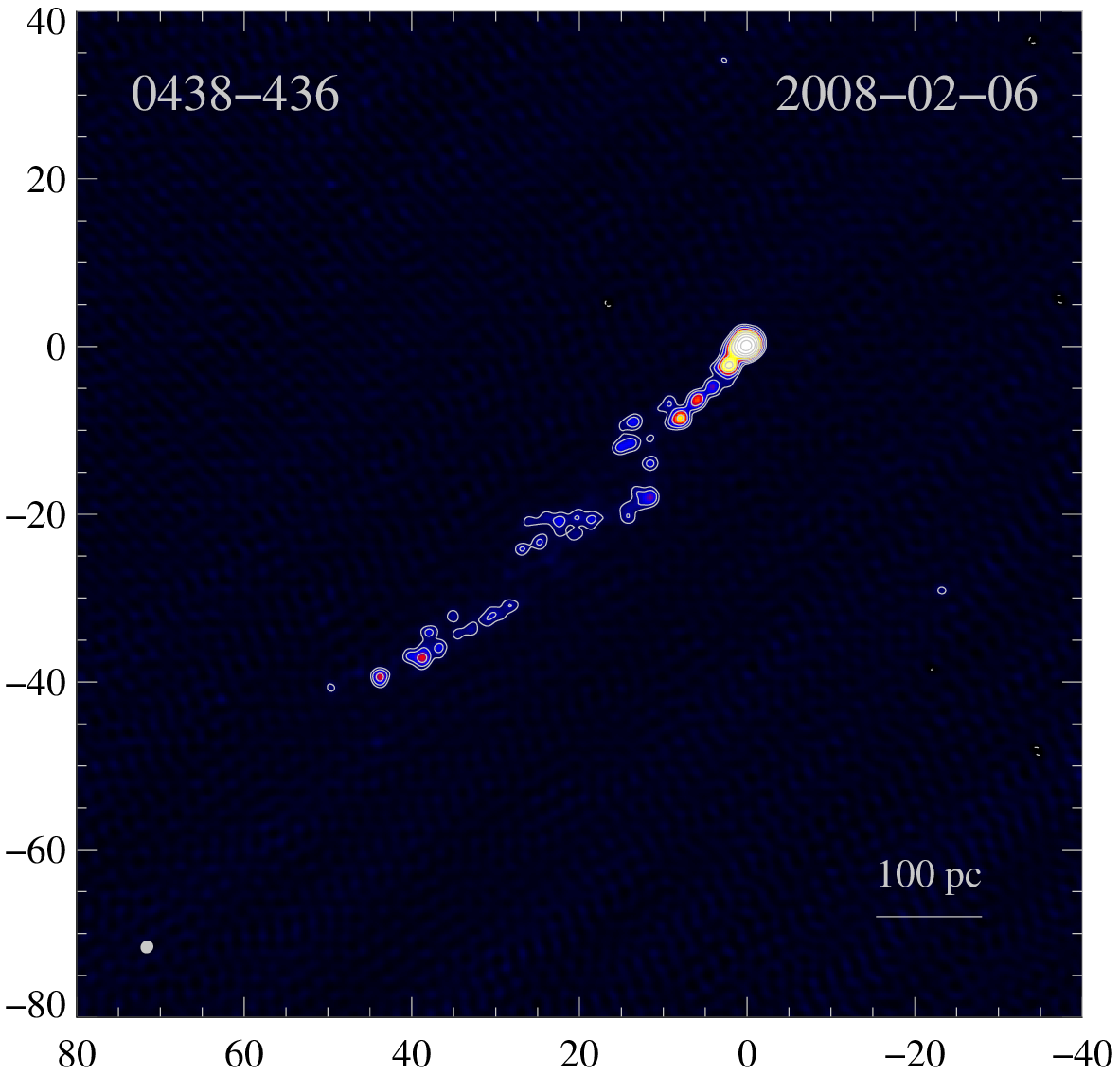}
\includegraphics[width=0.28\textwidth]{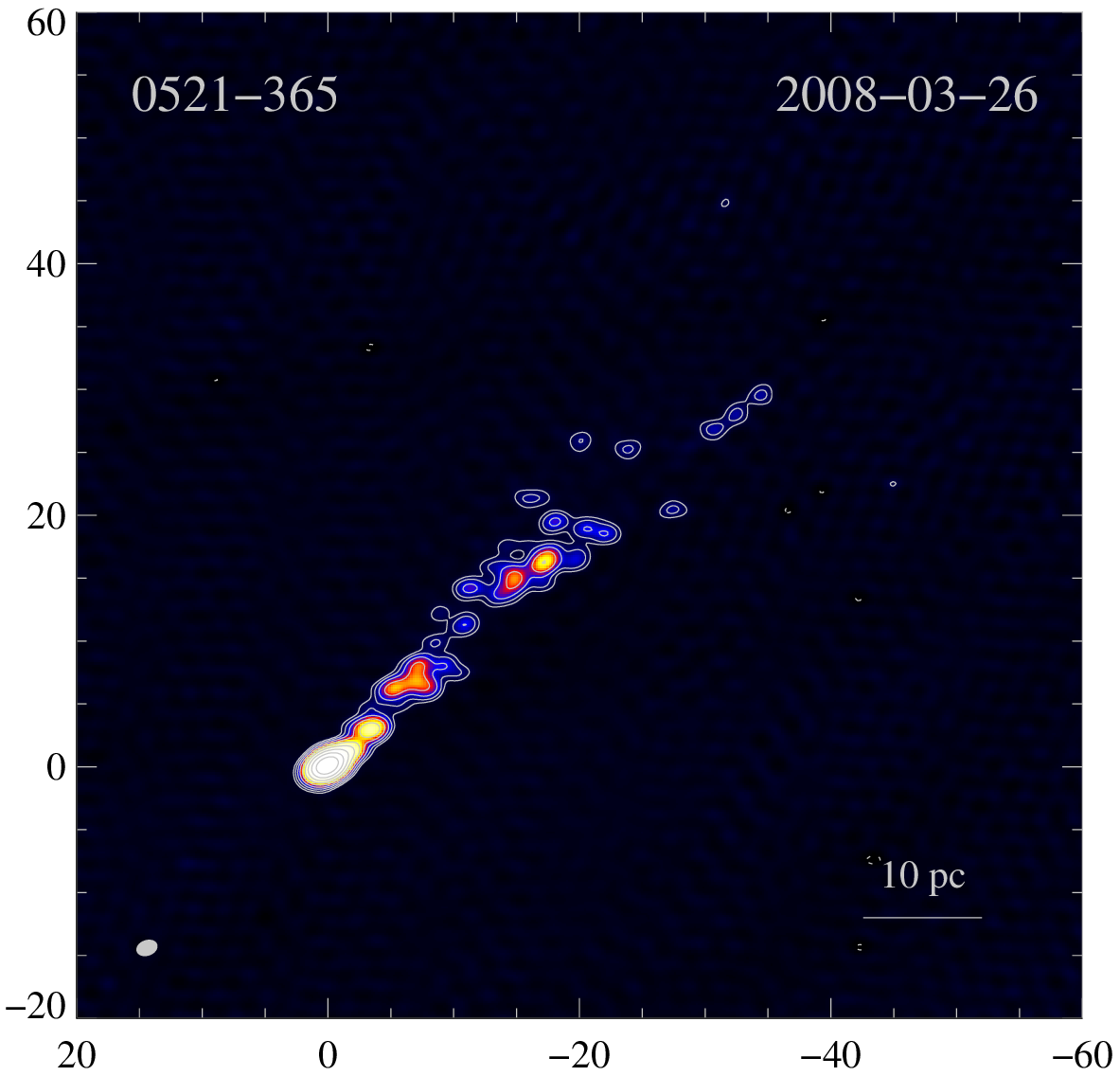}
\includegraphics[width=0.28\textwidth]{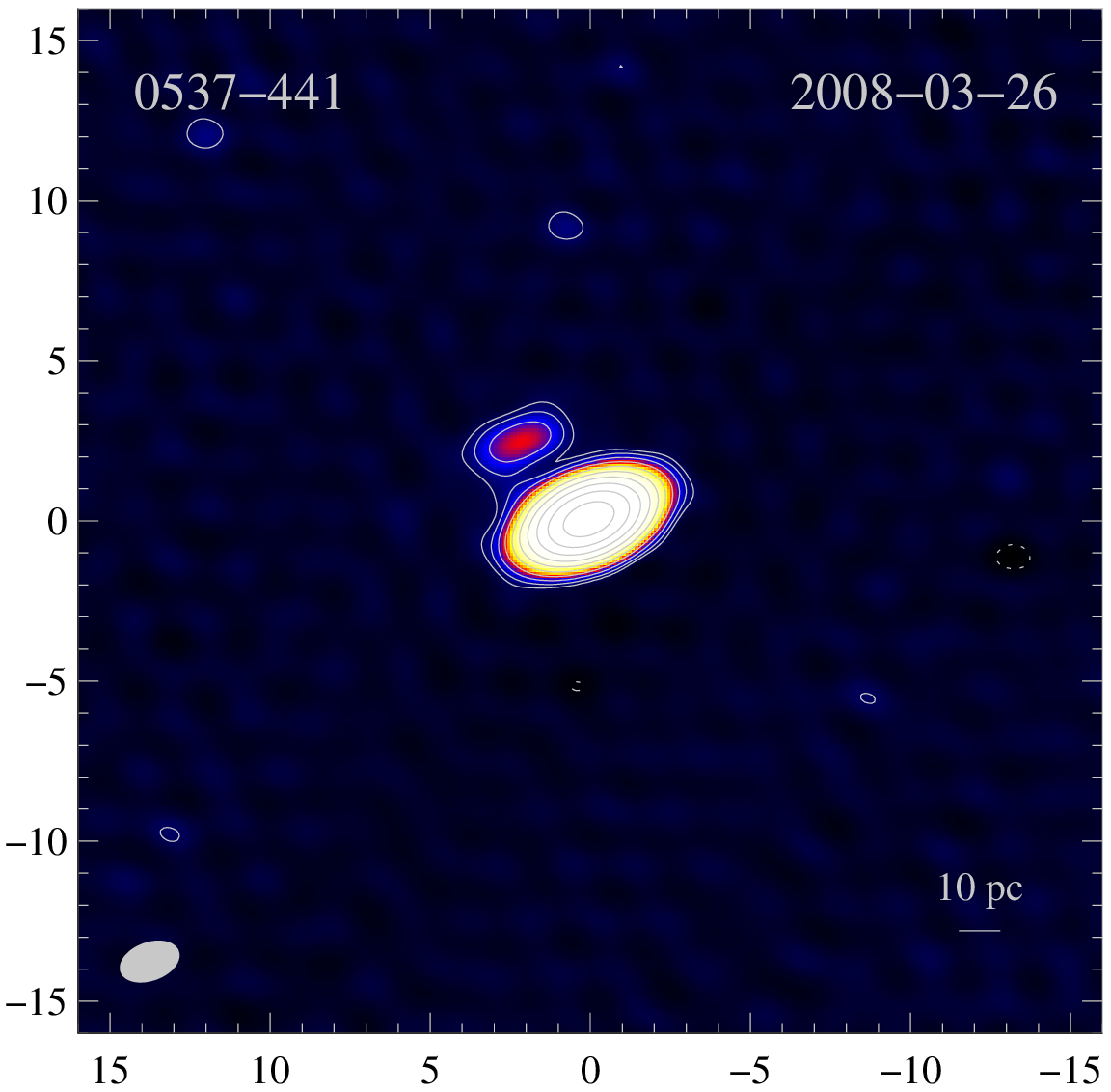}
\includegraphics[width=0.28\textwidth]{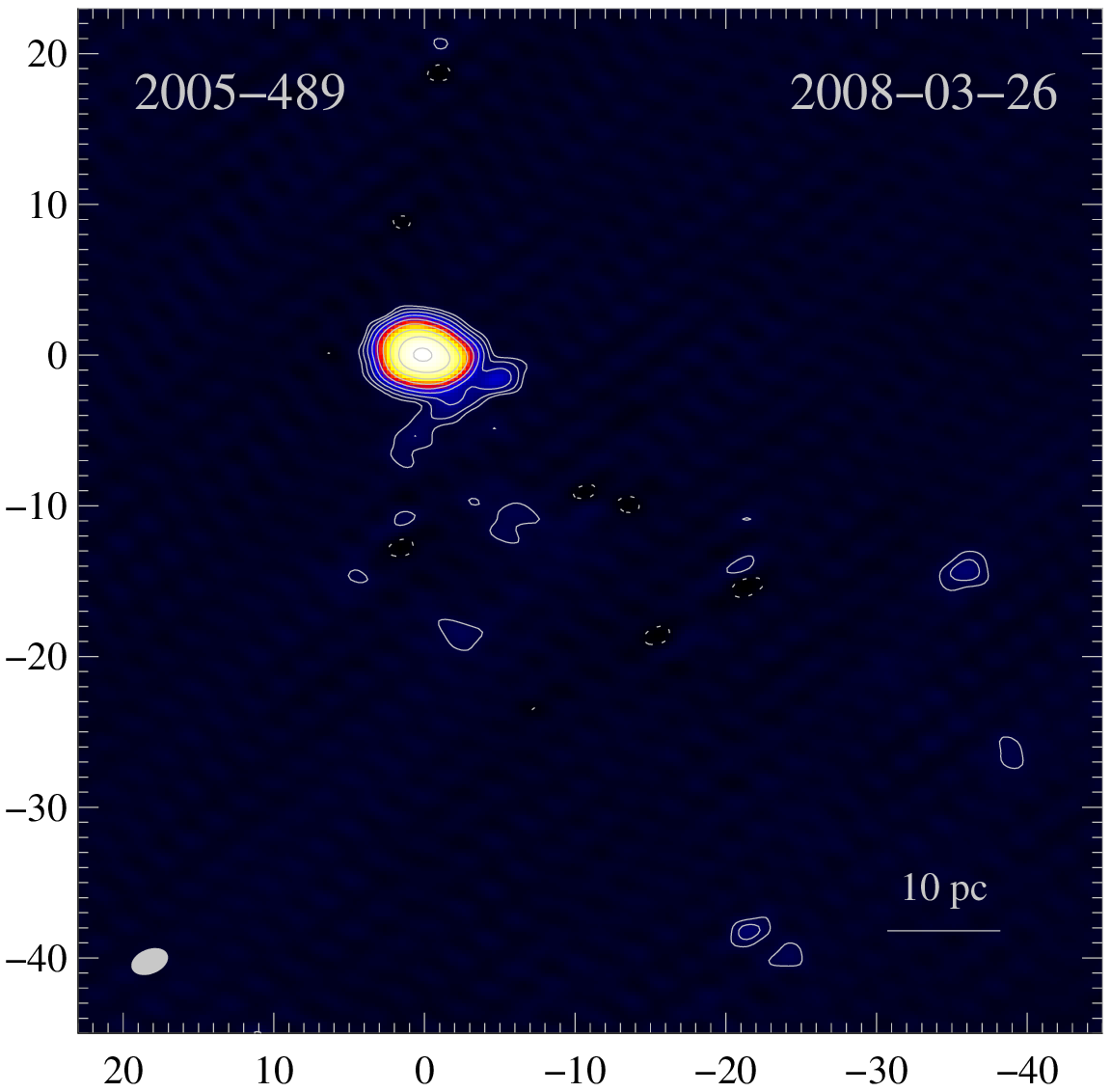}
\includegraphics[width=0.28\textwidth]{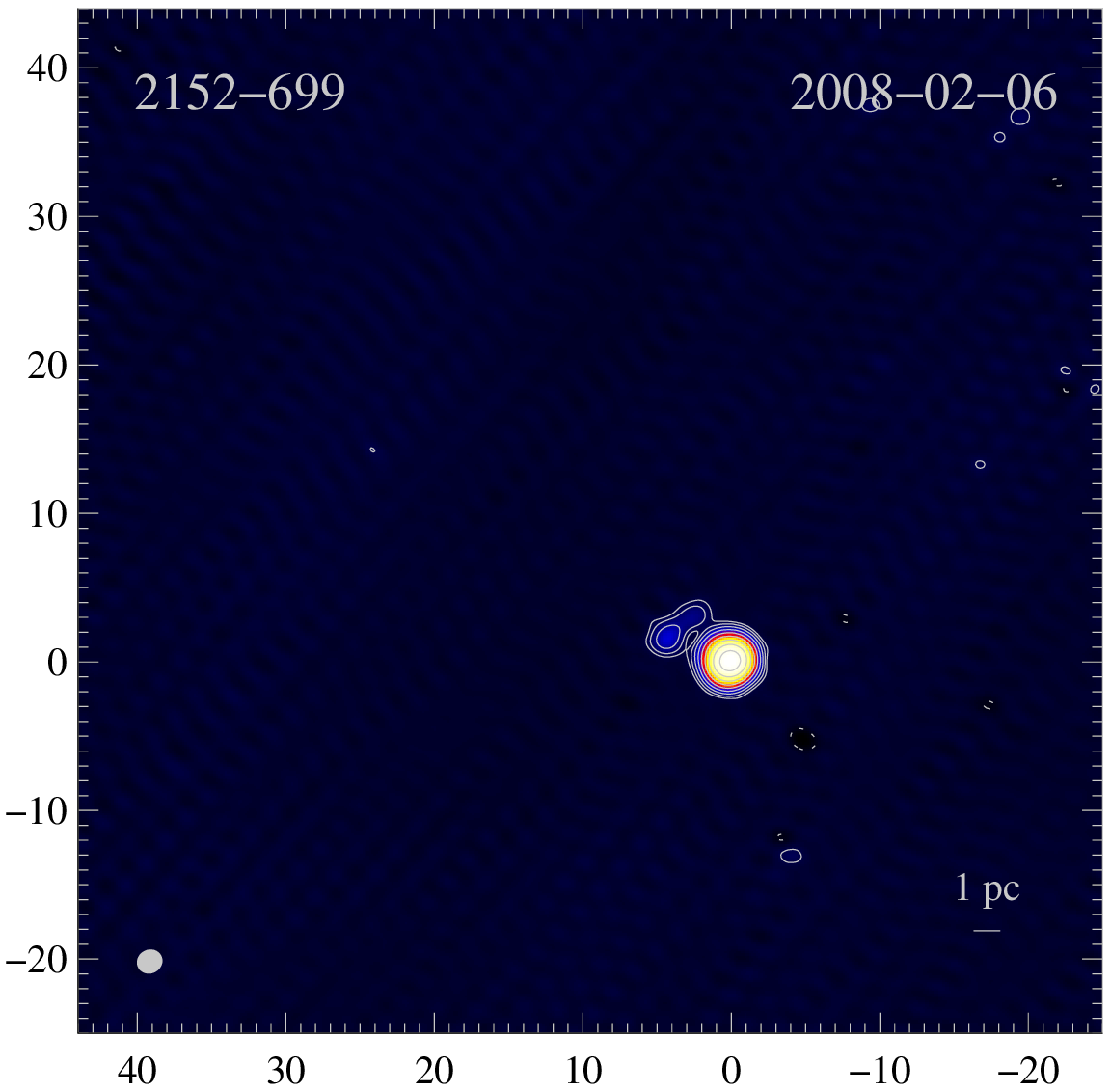}
\caption{Same sources as in Fig.~\ref{plots_X} at 22\,GHz}
\label{plots_K}
\end{figure*}
%%%%%%%%%%%%%
\section{Results and Discussion}
Figure~\ref{plots_X} and~\ref{plots_K} show a selection of naturally weighted images of some sources of the initial TANAMI sample at 8.4\,GHz and 22\,GHz, respectively. The full width half maximum (FWHM) Gaussian restoring beam applied to the images is shown as an ellipse in the lower left of each panel. Each panel also shows a bar representing a linear scale of 1\,pc, 10\,pc, or 100\,pc depending on the source extent and distance.

Figure~\ref{z} shows the redshift distribution of the extended TANAMI sample (see Table I). Galaxies and BL\,Lacs are distributed at $z<0.4$ while for quasars the distribution peaks at $\sim 1.3$, with a maximum redshift of $z=2.86$. This resembles the distributions for bright $\gamma$-ray AGN seen by LAT after three months of operation~\cite{Abdo2009}. There is no significant difference in the redshift distribution of the two sub-samples. Most galaxies have not been detected by the LAT, and secondly, none of the two most distant sources (see Table I).
For further results concerning brightness temperature, luminosity and morphology see Ojha et al., 2010 (submitted to A\&A) and other contributions to the \lq\lq Fermi Symposium 2009" proceedings (Ojha et al. 2009, submitted; B\"ock et al. 2009, submitted).

\section{Conclusion}
The LBA and additional telescopes in South Africa, Antarctica, and Chile provide high quality images of extragalactic jets at 8.4\,GHz and 22\,GHz that can be used to study physical properties of blazars. In many cases our images represent a substantial improvement on published work.

Our initial images and results presented by Ojha et al. will be corroborated by the further analysis of the 22\,GHz as well as multi-epoch data to consider spectral and kinematic information. Of particular interest will be TANAMI epochs observed since the launch of \textit{Fermi} providing near-simultaneous data at radio, $\gamma$-ray as well as at intermediate wavelengths from coordinated observations with other instruments.
\bigskip % extra skip inserted
\begin{acknowledgments}
This research has been partially funded by the Bundesministerium f\"ur Wirtschaft und Technologie under Deutsches Zentrum f\"ur Luft- und Raumfahrt grant number 50OR0808. The Long Baseline Array is part of the Australia Telescope which is funded by the Commonwealth of Australia for operation as a National Facility managed by CSIRO.  This research has made use of the NASA/IPAC Extragalactic Database (NED) which is operated by the Jet Propulsion Laboratory, California Institute of Technology, under contract with the National Aeronautics and Space Administration.
\end{acknowledgments}

\bigskip % extra skip inserted


\begin{thebibliography}{99} % Use for 10-99 references

\bibitem{Abdo2009}
Abdo, A.~A., et al., 2009, ApJ 700, 597

\bibitem{Deller2007}
Deller, A.~T., Tingay, S.~J., Bailes, M., \& West, C.\ 2007, PASP, 119, 318 

\bibitem{Dondi1995}
Dondi, L., Ghisellini, G., 1995, MNRAS 273, 583

\bibitem{Hartman1992} Hartman, R.~C., et al.\ 
1992, ApJ, 385, L1 

\bibitem{Healey2008}
Healey, S.~E., et al., 2008, ApJS, 175, 97-104

\bibitem{Lister2009}
Lister, M.~L., et al., 2009, AJ, 137, 3718 

\bibitem{Ojha2005}
Ojha, R., Fey, A.~L., Charlot, P., et al., 2005, AJ 130, 2529

\bibitem{Shepherd1997}
Shepherd, M.~C., 1997, Astronomical Data Analysis Software and Systems VI, 125, 77

\bibitem{Stickel1994}
Stickel, M., Meisenheimer, K., Kuehr, H., 1994, A\&AS 105, 211

\bibitem{Veron2006}
Veron-Cetty, M.~P., Veron, P., 2006, A\&A, 455, 773 


\end{thebibliography}
\end{document}